Overcoming Barriers to Engagement with Educational Video Games for Self-Directed Learning:

A Mixed-Methods Case Study

Angela Y. He[a]

[a]Oakton High School, Sutton Road 2900, Vienna, VA 22181-2900, United States. Corresponding author. E-mail address: heangela.arts@gmail.com

This research did not receive any specific grant from funding agencies in the public, commercial, or not-for-profit sectors.

Abstract

Research has established increased engagement and positive behavioral, attitudinal, and learning outcomes from educational games. Although engagement begets these benefits, there is a lack of research on how students engage with educational games, especially when self-directed. Additionally, research on the effects of engagement on performance is conflicting. This study aimed to identify barriers to students' self-directed engagement with two anatomy educational games, methods to overcome identified barriers, and the impact of engagement on learning performance. Employing the within-case and cross-case approach and triangulation of data, four educational-game specific barriers emerged (from most common to least common): 1) negative perceptions of educational games, 2) incompatible audience, 3) incompatible difficulty, and 4) price. Path analysis found that engagement may have insignificant, positive effects on performance. These findings indicate that students' self-directed engagement with educational games is impeded by unique barriers and engagement may not be vital for performance. Suggestions for game design and marketing emerged to help developers, teachers, and researchers overcome identified barriers.

Overcoming Barriers to Self-Directed Student Engagement with Educational Games: A Mixed-Methods Case Study

## 1. Introduction

In the past decades, the impacts of and relationships between educational video games (henceforth "games"), self-directed learning, and engagement have all received increasing academic attention, with relatively sparse but mostly positive results from educational games and more concurring substantiation supporting the latter two concepts (Boyle et al., 2016; Habgood, 2007; Hainey et al., 2016; Rashid & Asghar, 2016; Squire, 2008). Typically, educational games aim to teach through interaction and story, with learning content conforming to a standardized curriculum (Swain, 2007). With educational games being increasingly incorporated into a variety of learning environments, e.g. classrooms, government, finance, healthcare, hospitality, science and technology (Guillén-Nieto & Aleson-Carbonell, 2012), their potential seems limitless. In fact, educational games have potential to be utilized for self-directed learning, following the likes of massive open online courses (MOOCs), blogs, wikis, podcasts, and other self-directed learning technologies (Liu, Kalk, Kinney, & Orr, 2010). Self-directed learning can be broadly defined as learning ultimately initiated and managed by the learner, with or without external help (Knowles, 1989). Utilization of games for self-directed learning is possible due to increased student familiarity with and use of technology (Rashid & Asghar, 2016), and the flexibility and capabilities of game design to fulfill the Technology Acceptance Model (TAM), a prominent measurement for system's acceptance, and self-determination needs (Davis, 1993; Przybylski, Rigby, & Ryan, 2010). If used for self-directed learning, educational games could be more advantageous than other self-directed learning technologies by, in addition to knowledge acquisition and intrinsically motivated use, causing positive changes in students' cognitive abilities, attitude, and behavior (Boyle et al. 2016; Dobrowolski et al., 2015). However, according to the TAM, the user must first be willing to engage with a technology in order to fully utilize it (Davis, 1993). Therefore, self-directed engagement is pertinent to self-directed learning with educational games. Additionally, since whether or not a user is intrinsically motivated upon using a technology affects the user's derived benefits (Martens, Gulikers, & Bastiaens, 2004), self-directed engagement is relevant to non-autonomous usages of educational games. Thus, it is important to identify barriers to self-directed student engagement that prevent educational games' straightforward adoption.

*1.1. Games for self-directed learning*

According to recent literature reviews (Boyle et al., 2016; Hainey et al., 2016), existing educational gaming research has consistently established that game-based learning has positive effects on learning performance, as well as cognitive abilities such as critical thinking and visual acuity, motivation, attitudes, and engagement. One reason, besides the intrinsic design of games, could be increased familiarity with technology. While said familiarity seems to vary depending on demographics and geographics (Bulfin, Henderson, Johnson, & Selwyn, 2014; Shapiro et al., 2017), a growing number of authors (Becker, 2000; Luckin et al. 2009; Palfrey & Gasser, 2013; Tapscott, 2009) believe that a growth of and familiarity with information and communications technology in new generations lend itself to greater engagement with instructional technologies. Indeed, previous research demonstrates that entertaining educational games have stronger positive effects on student performance and enjoyment than traditional teaching methods (Ibrahim, Yusoff, Omar, & Jaafar, 2011; Ke, 2008; Suh, S. Kim, & N. Kim, 2010).

Previous studies indicate enjoyment improves performance and intention of use (Giannakos, 2013; Lee, Cheung, & Chen, 2005; Venkatesh, 2000). Increased intention to use and learning implicates potential for enjoyable educational games to be used for self-directed learning. Games are also better able to satisfy the three psychological needs for self-determination - competence, relatedness, and autonomy - (Przybylski, Rigby, & Ryan, 2010) and offer personalized rewards than other instructional technology (Schell, 2014), which would both facilitate self-directed learning. Additionally, because students, controlled for demographics, have greater familiarity with and usage of games over instructional technologies (Jacobsen & Forste, 2011), they may have greater willingness to use games for self-directed learning. In fact, Rashid and Asghar (2016) found that the usage of technology shares a positive relationship with self-directed learning. Unfortunately, within Rashid and Asghar's analysis, games showed a negligible effect on self-directed learning. Given the plausibility of educational games as self-directed learning tools, the evidence saying otherwise, and the increasing academic interest in self-directed learning, there is a need for illumination on self-directed learning via educational games.

This study aimed to address this gap by examining the first step of self-directed learning with technology: self-directed engagement.

*1.2. Self-directed engagement*

Although engagement is often described broadly and disparately, previous research (e.g., Banyte & Gadeikiene, 2015; Fredricks, Filsecker, & Lawson, 2016; Rashid & Asghar, 2016) most commonly agrees that engagement is a three-dimensional construct consisting of engaged behavior, affect, and cognition. Thus, an engaged individual would dedicate time, effort, and attention (behavior); have positive perceptions, beliefs, and intentions (affect); and use self-regulation and strategies (cognition) toward an activity (Fredricks, Filsecker, & Lawson, 2016). Self-directed engagement is engagement that, while it can be externally influenced, is not forced upon or mandatory to the engaging individual. For example, engaging beyond the classroom or in the classroom without it being mandatory is self-directed engagement.

      Self-directed engagement is important. For educational games to be used for self-directed learning, or successfully penetrate the commercial game market, students must first engage with educational games of their own volition. Indeed, aforementioned commercial game market penetration is being given increasing business consideration (Entertainment Software Association, 2016), thus raising the importance of self-directed engagement. While consumer research usually assumes engagement to be self-directed and therefore does not need to distinguish between engagement and self-directed engagement (e.g., Kim & Ko, 2012), educational gaming research often interprets engagement as being dictated by authorities, since the researcher or teacher typically determines if students will use a teaching method or not (e.g., Annetta, Minogue, Holmes, & Cheng, 2009; Papastergiou, 2009; Giannakos, 2013). As Martens, Gulikers, & Bastiaens (2004) put it: "Game playing has a strong resemblance with intrinsically motivated behaviour. In both cases the perception of 'fun' is crucial and too much external control is detrimental" (p. 370); thus, compulsory game playing results could potentially be different from those of voluntary game playing. To fill this gap in the education literature, this study will focus on barriers unique to self-directed engagement with educational games, rather than frameworks, heuristics, or other popular approaches in gaming research. Barriers analysis is used because research on nascent technologies often encounter numerous barriers to their

adoption (e.g., An & Reigeluth, 2011; Franklin & Molebash, 2007), foreshadowing that there might be barriers to autonomous student adoption of educational games, which must be overcome first and foremost. Due to existing, extensive research on barriers to self-directed learning with technology (Kormos & Csizér, 2013; Lai, 2015; Venkatesh, Croteau, & Rabah, 2014), commonly agreed upon barriers such as self-efficacy, perceived ease of use, perceived usefulness, external environment, and time management are not addressed directly—rather, specific barriers, solely applicable to educational games, are reported.

*1.3. Student perceptions of educational games*

Exploring potential barriers to engagement is doubly crucial when considering the gap in research on student perceptions of educational games. Although student perceptions of technology whose use is intrinsically motivated are generally positive, these technologies are mainly entertainment (Gurung & Rutledge, 2014). Thus, when games have an educational aspect, students' generally positive attitude and intrinsic motivation to play may differ. Indeed, Bourgonjon et al. (2010) found that students' beliefs of the learning opportunities within games positively impacted the students' preference to use games in the classroom. Although entertainment gaming researchers are not students, they are also invested in technology, and also insinuate a difference by not addressing educational games when researching game genres (Lee et al., 2007; Elliott, Golub, Ream, & Dunlap, 2012). Wolf (2001), who did acknowledge educational games when outlining game genres, states that "the degree to which [educational games] can be considered games varies greatly" (p. 124). Differences in student perception of a learning environment will indirectly impact student engagement by directly impacting intrinsic motivation (Beaudry & Pinsonneault, 2010; Driscoll, 2005; Patrick, Ryan, & Kaplan, 2007).

Thus, engagement with poorly perceived educational games would be deficient and might miss out on found benefits of intrinsically motivated engagement, such as curiosity, interest, exploration, experimentation, positive affect, and satisfaction (Habgood & Ainsworth, 2011; Hu & Hui, 2012; Hyland & Kranzow, 2012; Martens, Gulikers, & Bastiaens, 2004). Intrinsically motivated use is especially pertinent to educational games, since resistance threatens active participation which is fundamental to game-based learning (Squire, 2008). Because the education literature is deeply interested in intrinsic and extrinsic motivation, and research indicates intrinsic motivation leads to self-directed engagement (Banyte & Gadeikiene, 2015; Fredricks, Filsecker, & Lawson, 2016), and perceptions have direct impacts on emotional engagement itself, investigating student perception is important to this study's aim and the education community.

  While articles have been made in wholly understanding teacher perceptions and attitudes (Bourgonjon et al., 2013; Egenfeldt-Nielsen, 2004; Schifter & Ketelhut, 2009), as well as parental perceptions of educational games (Bourgonjon et al., 2011), much less is known about student perceptions of educational games. There is little existing literature on the topic, and they are restrained by quantitative or close-ended measurements which do not take into account other contexts for educational game usage, such as out-of-classroom settings (Bourgonjon, Valcke, Soetaert, & Schellens, 2010; Papastergiou, 2009). Additionally, a number of them assess student attitude towards the intervention game, not educational games in general (Giannakos, 2013; Papastergiou, 2009). A broader approach must be taken; as Bourgonjon et al. (2011) wrote: "a key issue is whether researchers adopt a sufficiently broad approach when studying the key actors in an instructional setting" (p. 1434). This deficiency in understanding students' over teachers' or parents' perceptions is especially important, considering that the students, not the teachers or parents, are the users of instructional technology and thus must be motivated to use it. Therefore, to fill this gap and fully understand barriers to students' self-directed engagement with educational games, students' perceptions must be considered with an open-ended approach.

*1.4. Importance of engagement to learning*

When studying a topic, a researcher must ask: is this topic important to the field? Therefore, as engagement is being studied, the importance of engagement to educational games must be analyzed. The research on the relationship between engagement and learning performance in educational games are conflicting. With data on 129 participants, Annetta, Minogue, Holmes, and Cheng (2009) found that engagement with their educational game intervention did not affect performance. However, their findings did not quantify engagement through an academically validated measurement, instead assuming that the intervention was more engaging than the control (traditional classroom teaching). In contrast, Blasco-Arcas, Buil, Hernández-Ortega, and Sese (2013) only used data measurements supported by literature and found that increased engagement with a technology improves learning performance. However, the technology in question is a clicker, which is not as complex as educational games. Other articles (e.g., Brown & Cairns, 2004) have hypothesized positive effects from engagement with games on learning performance among others, such as intrinsic motivation and enjoyment. A reason for the debate on engagement could be that confounding variables skewed the results of the articles in this section; for example, some articles did not measure demographics, which could have confounded the results. Clearly, more research on the relationship between engagement, performance, and potential confounders such as demographics is needed to ascertain the truth on engagement with games.

*1.5. Designing game for engagement*

Effective game elements promoting engagement and learning is an additional key area for academic exploration, as found in literature reviews from Boyle et al. (2016) and Schmid et al. (2014). While, contrary to these findings, there is a plentitude of game design literature on engaging game elements (Pinelle, Wong, & Stach, 2008; Rigby & Ryan, 2011; Schell, 2014; Sweetser & Wyeth, 2005), there could be more exploration in engaging game elements for educational games, specifically. This study is suited to contribute to this topic since, when analyzing student decisions in the engagement process, game design needs will inevitably crop up as either barriers or contributors to engagement. As it is currently, the gaming literature most agrees on interaction and multimedia as highly engaging game elements.

Researchers consented that interaction was conducive to all levels of engagement with games (Brown & Cairns, 2004; Chang, Liang, Chou, & Lin, 2017; Papastergiou, 2009). Some researchers find that only meaningful interactions, contributing to player agency and autonomy, are beneficial to engagement because they induce a sense of purpose and thus increased motivation (Mack & Nielsen, 1994; Przybylski, Rigby, & Ryan, 2010). Other researchers claim any interaction as contributory to engagement since, meaningful or not, it takes up germane cognitive load thus increasing cognitive engagement (Bouvier, Lavoué, Sehaba, and George 2013; Wiebe, Lamb, Hardy, and Sharek, 2014). However, these claims seem unsubstantiated when compared to the empirical evidence for meaningful interaction. Nevertheless, most researchers consent that some form of interaction and engagement are positively related.

To transition from low to high levels of engagement, researchers suggested emphasizing multimedia in games. For instance, Brown and Cairns's (2004) grounded theory of immersion found that the absence of atmospheric visuals and audio was a barrier to immersion. Likewise, in Papastergiou's (2009) more statistically significant study, students demanded for more multimedia in the intervention applications because multimedia helped retain interest. Another example is a finding from Sanders and Cairns (2010) that, when players were asked to keep track of time spent playing a game, the addition of 'likeable' music caused bigger underestimations. Seeing as losing track of time is a characteristic of deep engagement (Brown & Cairns, 2004; Ermi & Mäyrä, 2005), the results could imply that the music increased engagement. Moreover, Chen and Sun (2012) found that dynamic multimedia, e.g. video and animation, versus static multimedia increased affective engagement. More recently, Chang, Liang, Chou, and Lin (2017) claimed that the incorporation of effective multimedia, i.e. high quality visuals and audio of a cognitively manageable quantity, in a learning game increased concentration, positive affect, and germane cognitive load. Therefore, certain multimedia can enhance engagement.

Consequently, in developing two versions of an anatomy educational game for this study, both contained high quality music, and the more engaging version contained more moving visuals and meaningful interaction.

*1.6. Purpose of study*

The current study attempted to help fill the aforementioned gaps in educational gaming research by exploring self-directed engagement in the two intervention games, and by focusing on the following research questions: (1) How can barriers unique to self-directed engagement educational games be overcome? (2) Does increased engagement with an educational game improve performance?

## 2. Methods

*2.1. Research design*

Because holistic data from multiple students was needed to find common barriers on a relatively untouched topic (self-directed engagement with educational games), an exploratory multiple case study design was employed (Yin, 2013). One case was composed of one participant playing an intervention game. The case study compared two anatomy educational games. As one game was designed to be more engaging than the other, differences in data could be attributed to the differences between the games' designs. In accordance with Cohen, Manion, and Morrison's (2013) advice, mixed methods were used to produce the most comprehensive, convincing evidence for instructional technology. Data were triangulated using observations, interviews, and documents. Documents, in the form of three different questionnaires, acted as quantitative measurements, measuring within-group pretest-posttest performance scores, engagement, and demographics.

*2.2. Participants*

Ten students (4 males and 6 females) participated in this study. The sample was representative of their population, juniors and senior students from a high school in Fairfax County, Virginia, which was defined by the following characteristics: 17-18 years old, approximately equal proportions of males and females, mid-high income, and racially diverse. The school is more academically rigorous than the national average and located in a high-income area. Purposeful sampling, an extension of convenience sampling, was employed to leverage the researcher's in-depth familiarity with the population, thus selecting and ensuring a representative sample (Marshall, 1996). Age was chosen because older students are more likely to engage with games, seeing as the average game player is 35 (Entertainment Software Association, 2015). In addition to age, older students have increased access to products, especially technology (Jacobsen & Forste, 2011), thus are more likely candidates for self-directed engagement with instructional technology. The sample population also served to help fill a gap in game-based learning research on non-primary school students, as noted by Mayer et al. (2013).

Five students (3 female and 2 male) each were randomly assigned to one of two groups (group A and group B).

2.3. Materials

2.3.1. The two anatomy games

Two versions (game A and game B) of *Grey Plague*, a point and click adventure game developed by the researcher, were used in this study. Group A played game A; group B played game B. Game B is free-to-play online at https://zephyo.itch.io/grey-plague. *Grey Plague*'s purpose is to immerse while introducing basic concepts on human anatomy, physiology, and tuberculosis. The anatomy and physiology content conforms to the participants' high school Anatomy & Physiology curriculum. Tuberculosis content serves as a plot device and medical advanced-professional learning, i.e. learning not part of K12 education that facilitates professional success (Mayer et al., 2013).

The learning content of Grey Plague specifically teaches: (a) the location of vital organs (trachea, lungs, heart, diaphragm, intestines), (b) the parts of the lungs (trachea, bronchi, bronchiole, alveoli) and their functions, (c) the parts of the nervous system (central, peripheral) and their functions, (d) the parts of the brain (cerebrum, lobes, diencephalon, brain stem, cerebellum) and their functions, (e) the types of muscles (skeletal, smooth, cardiac), (f) the function of bones, (g) the definition and limitations of 3D organ printing, (h) the definition, infection process, and symptoms of tuberculosis.

The games' design purposefully adopted the following elements for general and educational game design (Annetta, 2010; Pinelle, Wong, & Stach, 2008; Rigby & Ryan, 2011; Schell, 2014): (a) rules and rewarding goals, (b) predictable controls, (c) atmospheric music and sound effects, (d) identity, (e) 3-dimensional characters, (f) intuitive user interface, (g) visual aids for learning content, as it was suggested by Khot, Quinlan, Norman, & Wainman (2013) to assist medical science education (Fig. 2).

The gameplay is mouse-only and conventional for the games' genres. For example, gameplay includes the conventional game mechanics of dialogue options, clicking to continue or fast forward typing effect of dialogue, and an inventory system, among others.

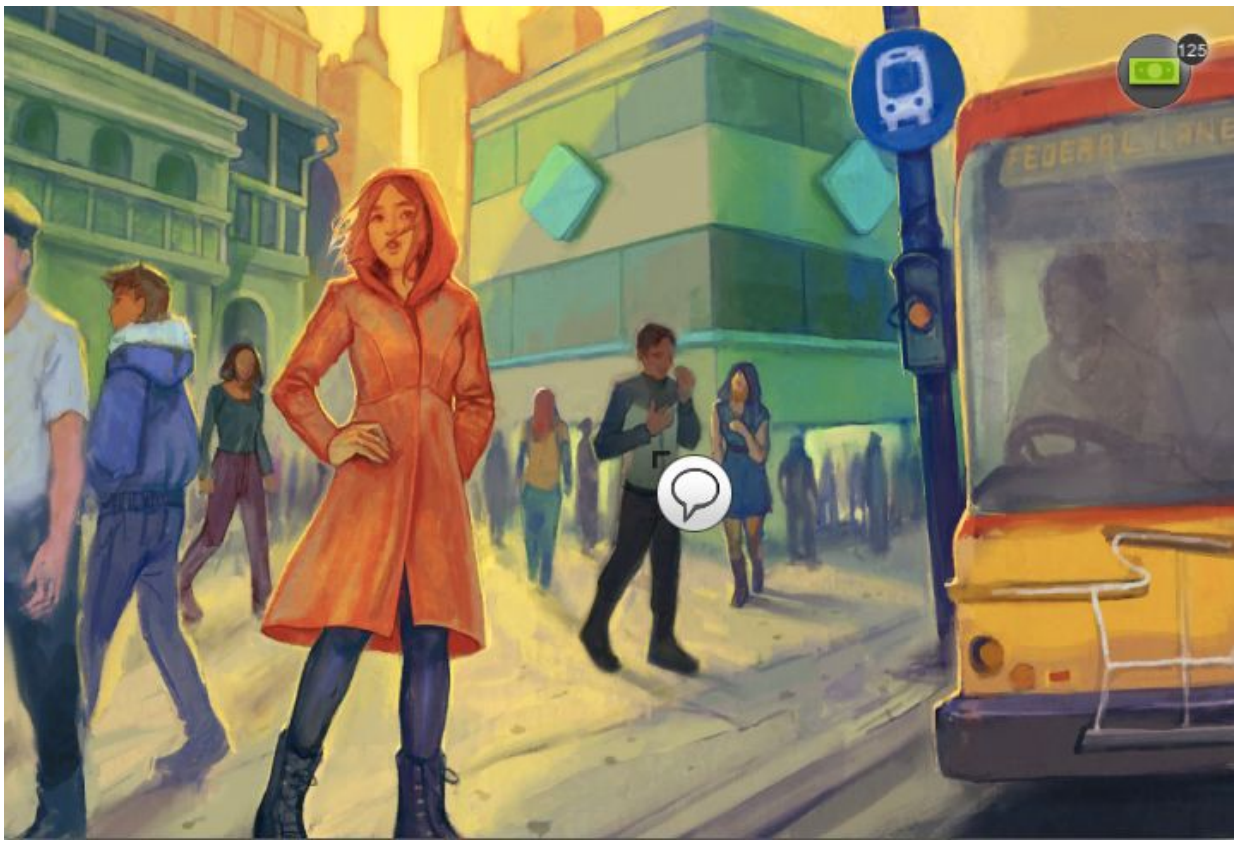

**Fig. 1.** Talking, examining, and entering actions are detected via user interface.

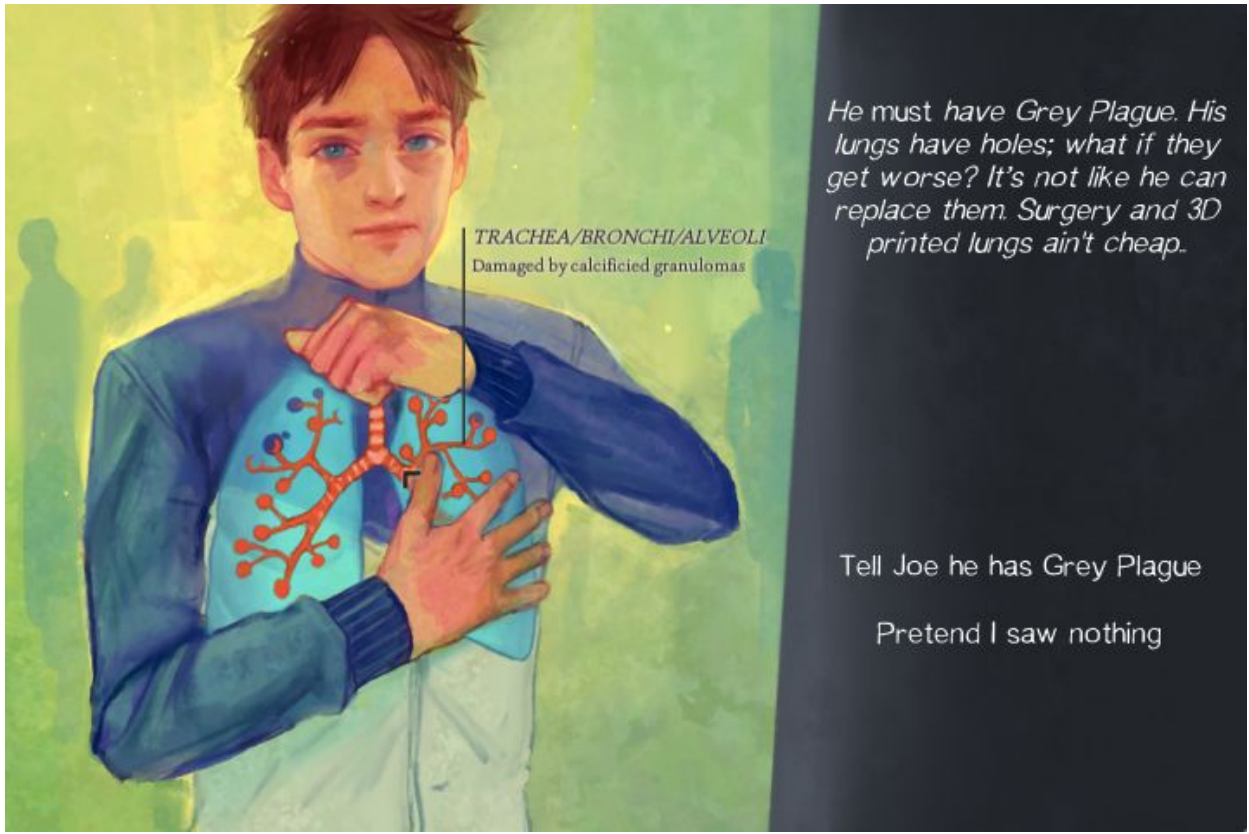

**Fig. 2.** Interactive diagrams give more information upon mouseover. Here, player faces moral dilemma, implemented for engagement.

The story, based off of Vogler's (2007) *Writer's Journey*, is as follows: It's 2070. A mysterious deadly disease has broken out. Dubbed 'Grey Plague' by physicians, the disease is suspected to be a mutation of tuberculosis, which was thought to have been eliminated years ago. The protagonist has the power to see others' internal anatomies and is ashamed of it. She gets a bloody note from her grandma—a call to adventure—which reveals her grandma has Grey Plague. Thus, she tries to transcend her anxiety of her power so she can save her Grandma and everyone else from Grey Plague.

*Grey Plague* was developed in Unity using C#, employing graphics built in Photoshop and audio sourced from Creative Commons.

*2.3.1.1. Differences between game A and B*

The difference between game A and B occurs while the protagonist's mother, a doctor, explains the infection process of tuberculosis to the protagonist. Participants took about 5-7 minutes playing this distinct section.

During this section, game A uses occasionally interactive, mainly static visuals accompanied by blandly-toned text to elucidate the infection process (Fig. 3). Thus, this section in game A could be classified as an unengaging visual novel, considering its emphasis on text, bland and specialized language, static imagery, and lack of interaction (Yin, Ring, & Bickmore, 2012). In this section in game B, the mother instead encourages the protagonist to imagine themselves as a tuberculosis bacterium. The protagonist then imagines and controls said bacterium, guiding it through the infection process (Fig. 4). This section in game B could be categorized as a strategy or action game, seeing as it incorporates louder, faster music; more interaction; constantly dynamic visuals; mazes; and a possibility of dying via contact with white blood cells. Thus, it engages the player more by increasing germane cognitive load and motivation to survive, or "win" (see also 1.3).

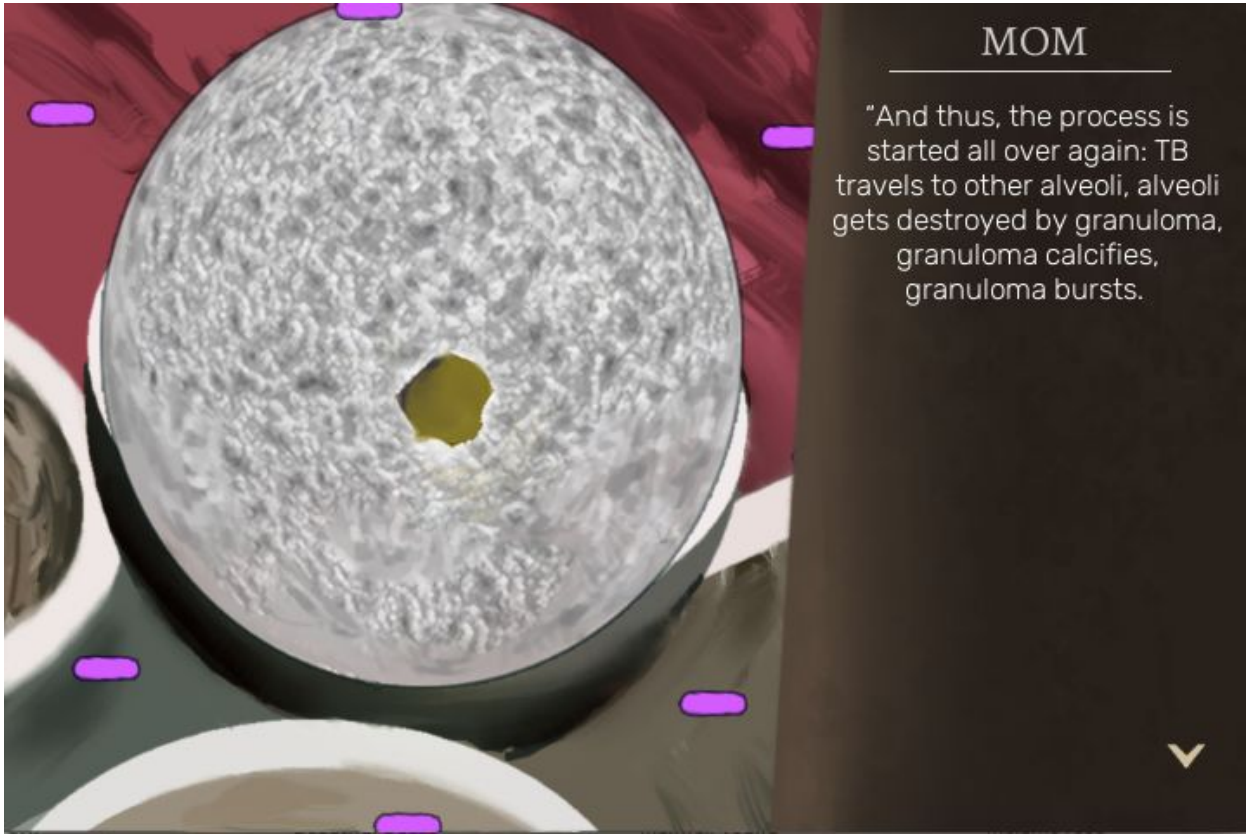

**Fig. 3.** The last sentence, accompanied by short-term animation, before the distinct section of game A ends.

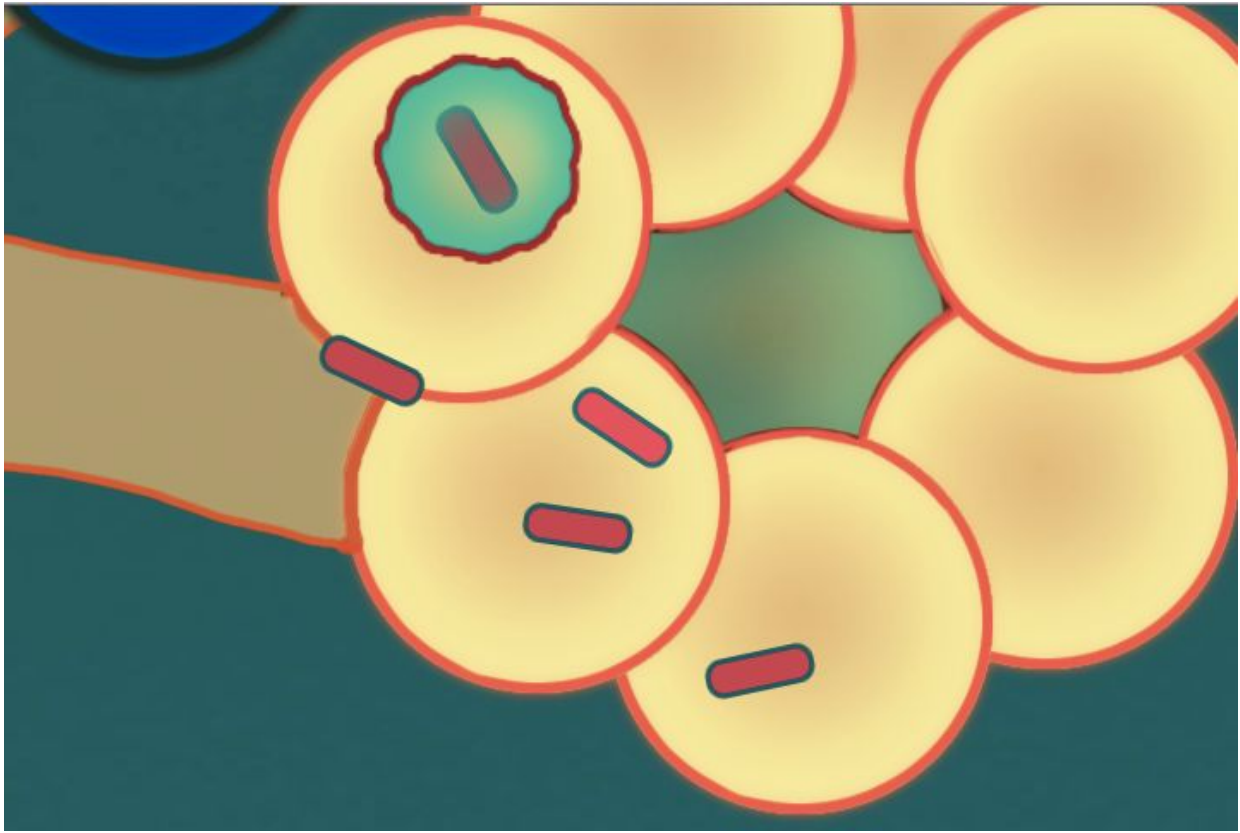

**Fig. 4.** The last level of the distinct section of game B. The player (lightest bacillus) struggles with white blood cells in an alveolus.

*2.3.2. The questionnaires*

Three questionnaires for document analysis were used to collect quantitative data. Questionnaires were hosted digitally, with Google Forms, as recommended by Habgood (2007) to conceal survey lengths and facilitate marking and data analysis.

The same test, used as the pretest and posttest, had 10 multiple choice questions, each with one correct answer, in a randomized order (Appendix A). Some questions assessed transfer of taught concepts by placing the concepts in realistic contexts. Questions were chosen from Seeley, Stephens, and Tate's *Anatomy and Physiology* editions 6 and 7 (2003, 2004).

An engagement assessment for quantifying engagement with *Grey Plague* consisted of 25 questions using five-point Likert scales. Questions were selected from Jennett et al.'s (2008) Experiment 3 game immersion questionnaire. Seeing as immersion in a game is essentially a high level of engagement with the game (Brown & Cairns, 2004), the substitution seemed reasonable. Moreover, a questionnaire on immersion instead of engagement was used because little substantiated game engagement questionnaires were found: proposed self-report instruments that target engagement with games (e.g., Hamari et al., 2016; Wiebe, Lamb, Hardy, & Sharek, 2014) are less substantiated than Jennett et al.'s (2008) measurement, perhaps due to their recency. Selected questions were engagement-specific, addressing such factors as motivation, enjoyment, attention, effort, and other attitudinal, behavioral, and cognitive factors (Fredricks, Filsecker, & Lawson, 2016).

The demographics questionnaire consisted of five multiple choice questions and three 5-point likert scale questions determining gender, age, gaming habits, edutainment habits, desired features in entertainment media, and average non-AP science classes grade (Appendix D). The purpose of demographics data was to analyze if demographics were confounding variables in the relationship between engagement and performance or quantifiable barriers to engagement.

*2.4. Procedure*

Data for each participant were gathered on campus after classes. Each participant played on a MacBook Air with stereo sound through headphones. Each game took about 160 MB on the computer's hard drive.

Before playing the intervention, a pretest score was collected. Then, the participant was observed as they finished the game in approximately thirty minutes. Behaviors, body language, and facial expressions when interacting with the game and external environment were recorded. During this time, the researcher did not initiate communication with the participant.

Afterward, as feelings of engagement are most susceptible to memory decay, the engagement score was collected first (Gass & Mackey, 2016). Its Likert scales were treated as interval data, as suggested by Brown (2011), so that descriptive statistics could be applied. Then, interview data were collected before collecting posttest score to mitigate reactivity (Gass & Mackey, 2016).

The face-to-face semi-structured interviews were recorded with an audio device. Interview questions were modelled after SurveyMonkey's online example product feedback surveys (SurveyMonkey, n.d.). As suggested by Brenner (2006), the first few questions were descriptive, relatable, and broad to facilitate the interviewee's comfort with the questioning. Open-ended questions on knowledge, attitudes, and behavior towards *Grey Plague* and educational games in general were asked to get a comprehensive understanding of each case (Appendix B). After the interview concluded, posttest performance score was collected. Demographics were collected last to prevent respondent fatigue affecting previously collected data, which was more pertinent to the research questions.

*2.4.1. Data analysis*

Quantitative analysis was performed using SPSS 24.0 and a confidence level of 95%. The aim of this analysis was to investigate how engagement affected performance, how the groups affected engagement and performance, and how demographic affected engagement and performance.

Qualitative analysis, was performed via coding using NVivo 11.4.0, a qualitative analysis program. Creswell's (2007) and Yin's (2013) recommended exploratory case study methods—within-case and cross-case—were followed. Within-case analysis finds unique themes in each case; afterwards, cross-case analysis categorizes similarities and differences between within-case findings (Yin, 2013). Following Wong's (2008) guide to NVivo, an initial round of open coding to identify themes within each case was performed. 49 total descriptive codes were found. Categories were developed by repeatedly removing overlapping categories, creating associations between categories, and capturing the main concept of remaining categories. Through this coding process, four main barriers emerged, each with suggestions to overcome it.

### 3. Results and discussion

*3.1. Quantitative results*

The means of the dependent variables pretest, posttest, performance (post-pre), and engagement within the groups are shown in Table 1.

**Table 1**

Descriptive statistics for pretest, posttest, performance, and engagement by group

| Variable | Group A | | Group B | |
| --- | --- | --- | --- | --- |
| | M | SD | M | SD |
| Pretest[a] | 5.40 | 2.51 | 4.80 | 1.79 |
| Posttest[a] | 6.80 | 2.78 | 7.20 | 1.30 |
| Performance | 1.40 | 1.82 | 2.40 | 1.95 |

| Engagement[b] | 85.20 | 10.52 | 92.20 | 7.89 |

[a]The full score was 10.

[b]The full score was 115.

To ensure compliance of parametric assumptions, normality of the dependent variables was assumed with the Shapiro-Wilk test and homogeneity of variances was assumed with Levene's test.

Pearson product-moment correlations revealed a positive yet insignificant correlation between engagement and performance (r=0.27, n=10, p=0.45). Engagement in Group B had a stronger correlation with performance (r=0.29, n=5, p=0.64) than that of Group A (r=0.1, n=5, p=0.87). Answering RQ2, Statistical analyses indicated that increased engagement did not significantly improve learning performance, but as the level of engagement increases, potential to may increase. The nonsignificance of engagement on learning is corroborated by previous research finding similar results (Annetta, Minogue, Holmes, & Cheng, 2009) and conflicts other research finding significant impacts (Hsu, Tsai, & Wang, 2012; Huizenga, Admiraal, Akkerman, & Ten Dam, 2009). Interestingly, Hamari et al. (2016) found a significant, positive effect of engagement on learning, but none of immersion on learning, which implicates a significant distinction between engagement and immersion.

Analyses of interviews suggest that higher proportions of attention to story and multimedia elements over learning content may explain the insignificance. This reasoning is connects with findings from Martens, Gulikers, and Bastiaens (2004) that increased intrinsic motivation did not lead to increased performance, but rather increased attention and exploration of an instructional technology. If this reasoning is true, then it implicates that increased attention from increased engagement must be somehow directed towards learning, not play, to improve performance. Alternatively, insignificance may be due to an insufficient sample size or difference between games. A one-way MANOVA, which indicated no significant difference between groups on engagement or performance ($F(2, 8) = 0.79$, $p=0.49$), substantiates the supposition of difference inadequacy between games. Within the multivariate analysis, univariate tests showed a stronger effect on engagement ($F(1, 9) = 1.42$, $p = 0.27$) than on performance ($F(1, 9) = 0.70$, $p = 0.43$), signalling that conscious incorporation of more effective multimedia and meaningful interaction worked to improve engagement.

A dependent *t*-test indicated a statistically significant improvement in test scores following the game treatment across groups from 5.10±2.08 to 7.00±2.06 ($t(9)=-3.24$, $p=0.01$); an improvement of 1.90±1.85. The effect size ($d=1.90/1.85=-1.03$) was calculated to be between 'large' and 'very large' (Cohen, 1988; Sawilowsky, 2009). Pretest ($M1$) and posttest ($M_2$) variables showed a positive correlation ($r=0.59$, $n=10$, $p=0.07$), substantiating the uniformity of the treatment effect across individuals. *T*-test results indicate that game is of same caliber as previous research interventions (Giannakos, 2013; Habgood, 2007; Papastergiou, 2009) which find significant impacts on performance, thus ensuring results are not impacted by learning inefficacy of game A and B.

*3.1.1. Demographics*

Using one-way ANOVAs, control variables of age, gender, grade (self-reported average grade in non-AP, mandatory science classes), and edutainment playtime demonstrated no significant effect on the dependent variables, engagement and performance. Gender results are consistent with previous research regarding insignificant or small differences between gender on experience with an educational game (Bourgonjon et al. 2010; Bourgonjon et al. 2011; Ke, 2008; Papastergiou, 2009).

Experience with games showed potential as a barrier to engagement. Among participants, three categories emerged: gamers, occasional gamers, and non-gamers. Gamers played games for 4 or more hours on a "typical non-school day" and showed extensive knowledge of games by referring to specific games and specialized terms such as DLC, gifting, etc. Occasional gamers demonstrated less playtime and knowledge of games; non-gamers didn't play games at all and had little knowledge about games. The conceptualization of experience fulfilled the need to transcend solely using of time spent on games to measure experience with games, and instead take a more multifaceted approach (Bourgonjon et al., 2011). Categorization into gamers (n=4) and tentative gamers plus non-gamers (n=5) resulted in a low, insignificant p-value for engagement ($F(1,8)=1.61$, $p=0.24$) and a lower one for performance ($F(1,8)=2.84$, $p=0.13$). Despite the difference in performance, there was no significant difference in grades between gamer categories ($F(1,8)=0.44$, $p=0.53$).

Since the number of gamers versus others was so small (n<6), results were interpreted with great caution. Even though more gaming experience may decrease engagement and increase performance, the chance of outliers or confounders (e.g. other demographics, attitudinal factors, personal innovativeness) impacting results was high. Indeed, in alignment with this finding, some researchers found that more experience led to increased performance (Egenfeldt-Nielsen, 2007; Verheul & van Dijck, 2009). However, mediation researchers found that the effect of experience or skill on learning is partially mediated by engagement (Hamari et al., 2016; Nikken & Jansz, 2006). More research to ascertain this relationship is needed; thus, experience will not be qualified as a barrier in this paper. Findings suggest relationship between gaming experience, engagement, and performance is complex.

The features participants sought in entertainment media signalled an insignificant effect on engagement ($F(3,6)=2.32$, $p=0.17$) and none on performance ($F(3,6)=0.42$, $p=0.74$). Features were determined via this question:

> (Select all that apply.) When seeking entertainment media (movies, books, TV shows, games, comics, etc), I usually look for the following features:
> 
> ❏ Entertainment value
> 
> ❏ Educational value
> 
> ❏ Innovation/novelty

All participants sought "entertainment value," with those only seeking "entertainment value" having lower engagement than those who sought an additional feature. This result indicates that intention to use may be a barrier to engagement and confounding variable, thus should be controlled for in similar studies. Due to insignificance of this result, intention to use will not be averred as a barrier. This finding implicates that students' intention to use for entertainment is ubiquitous, therefore, educational games should be entertaining to ensure engagement.

*3.2. Qualitative results*

Four most common barriers, specific to educational games, arose from within-case and cross-case thematic analysis of the collected documents, interviews, and observations: (1) [1] negative reputation of educational games, (2) incompatibility with older audiences, (3) incompatible difficulty, (4) price.

---

[1] Pseudonyms are used to replace the names of participants.

*3.2.1. Barrier 1: negative reputation of educational games*

Eight participants had moderately negative perceptions and beliefs, one had a neutral, open-minded approach, and another had a positive, advocating view of educational games. Negative perceptions mostly consisted of negative generalizations, insinuating that the negativity is directed towards educational games as a whole:

Researcher: What would you most like to improve or change about edutainment in general?

- Katie[1], group A, non-gamer: Most of them aren't that engaging; they're just kind of unnecessary because they're really inefficient and take so long.
- Brenda, group B, gamer: Stop trying too hard. For example, Rosetta Stone tries to be fun, but it feels like a drill exercise.
- Hannah, group A and occasional gamer: I don't play a lot of educational games; I don't really know much; but, there are the games that instantly tell you, "hey, this is an educational game, this is going to be really boring, and all you're going to do is just learn about math."

The definition of reputation, in a business sense, is the distribution of cognitive representations that consumers hold about a product or organization (Grunig & Hung-Baesecke, 2015), with cognitive representations meaning ideas about relationships between objects and attributes. Because participants had similar ideas in associating certain objects ("educational games") with certain attributes (e.g., "aren't that engaging," "really boring," "super boring"), participants' perceptions are conclusively part of a broader negative reputation of educational games. Note that despite this negative reputation, most participants preferred learning via games over traditional teaching methods, indicating that educational games are viable for more engaging teaching. For instance, Aiden, group B and gamer, stated, "I would really like to see educational games as a means of educating people in a more engaging way versus a traditional class. I feel like if I sat through a powerpoint that was about the same info.. I definitely wouldn't have been interested in learning about it in own time." The two participants who did not hold negative perceptions held characteristics of lifelong learners, e.g. enjoyed learning and were intrinsically motivated to learn (For example, the neutral participant described herself as "a student who's interested in biological functions." When recalling experiences with Civilization, a popular history game, the positive participant said "I spent the same amount of time reading the Civilopedia as I did playing the actual game"). Six of the negative perceptions had an underpinning of skepticism: when discussing their initial impression of *Grey Plague*, these six participants felt "pretty skeptical about a game based around education" or "surprised at how well done it was." Participants did not hold similar perceptions in response to the question, "What would you most like to improve or change about games in general?" Entertainment games had an overall moderately positive reputation, aligning with Boyle, Connolly, and Hainey's (2011) review of entertainment games' high psychological appeal. Any negative feelings about entertainment games were directed at specific games and addressed their quality or business model, not their game design ("in Minecraft, there are a lot of weird glitches" or "Marvel Heroes hasn't been updated in a few years"). Thus, this barrier is unique to educational games.

Overcoming this barrier is important for educational gaming researchers and businesses alike. Student engagement, self-directed or not, will be indirectly impaired from widespread negative or skeptical perceptions of the learning environment via decreased motivation (Driscoll, 2005, Patrick, Ryan, & Kaplan, 2007). Although the effect of engagement on learning in educational games isn't ascertained (Annetta, Minogue, Holmes, & Cheng; Blasco-Arcas, Buil, Hernández-Ortega, & Sese, 2013), impaired engagement with technology will most definitely negatively affect motivation, positive emotion, attention, and satisfaction with the technology (Arnone, Small, Chauncey, & McKenna, 2011; Chang, Liang, Chou, & Lin, 2017; Driscoll, 2005; Hu & Hui, 2012; Hyland & Kranzow, 2012). This implicates that, if previous studies observing attitudes toward educational games did not control for perceptions, their results may have been confounded, and have possibly underestimated.

For businesses, their product's reputation is one of their most important assets (Fombrun, 1996). Bad reputation of a product has multiple consequences, including decreased financial performance, product satisfaction, competitive advantages, and consumer loyalty (Aakar, 1991; Rogerson, 1983; Lai, Chiu, Yang, & Pai, 2010), all of which imply decreased total self-directed engagement. If the bad reputation of a product becomes massively well-known, potential consumers will shun or stigmatize the product, even if they have never experienced the product or their only knowledge of the product is from others (Grunig & Hung-Baesecke, 2015).

Causation for this barrier varied. Some negative perceptions of educational games were due to sparse but poor past experiences with edutainment. The poorness of these experiences were attributed to unenjoyable design decisions. Other participants' negative perceptions were due to negative perceptions of the subjects taught in the game, derived from negative relevant classroom experiences or social influence. For instance, Brenda "took the [Anatomy & Physiology] class last year and thought it was boring," and Sophia, group A and non-gamer, remarked that anatomy "seems more intriguing than what I've heard from other people talking about it." This is congruous with Shapiro et al. (2017) findings that previous bad classroom experiences with certain subjects occasionally led to disinterest or negative perceptions of the subject and decreased relevant MOOC participation. Unlike MOOCs, games have a reputation for being entertaining (Boyle, Connolly, & Hainey, 2011). This reputation may have led to the underlying skepticism in some participants' perceptions. Students generally do not associate enjoyment with learning and vice versa, so participants were doubtful about the success of combining the two (Prensky, 2002). Brenda remarked, surprised, that "while [*Grey Plague*] is educational, it's not super boring." This reasoning for skepticism is in alignment with research on priming, a psychological effect which finds that associations to a stimulus (in this case, entertainment games and past learning experiences) can be salient for the response to a related stimulus (educational games) (Herr, Sherman, & Fazio, 1982; Kolb & Whishaw, 2003).

Research by Ibrahim, Yusoff, Omar, and Jaafar (2011) contradicts these findings about negative reputation; their study indicated that 81% of their sample were "very interested in using games for learning in the future" (p. 213), indicating a positive reputation. However, their sample consisted of undergraduate students in Malaysia who were undertaking the course *Introduction to Programming*. Perhaps age or geographics influence perceptions of educational games, or students interested in programming are more partial to educational games. One commonality between this study and Irahim et al.'s study is that both measure perceptions after the educational game intervention. In hindsight, the intervention, as a representative of educational games, could influence student perceptions. Student perceptions before any intervention, across a variety of ages and interests, should be measured to clarify this confusion.

With the proper steps, this barrier is rather easily overcome because students' perceptions are not cemented. Students have not been exposed to much edutainment - seven participants self-reportedly using 0-1 hours of edutainment a week. Some participants had no exposure; Edward, group A and gamer, responded to a question on edutainment with "Yeah, this is my first one, so I don't really know." After playing *Grey Plague*, participants' negative or skeptical perceptions were mitigated by their enjoyable experiences. This straightforward overcoming can be explained by a lack of experience, which causes little priming or anchoring effects. Anchoring, described by Strack and Mussweiler (1997), is a cognitive bias describing the tendency for an individual's judgement of a task to center around their first experience, or anchor, of that task. It is difficult to avoid, even if anchors defy logic (Strack & Mussweiler, 1997). To establish positive impressions - anchors - and overcome this barrier, students should be exposed more to enjoyable educational games.

*3.2.1.1. Solutions to barrier 1*

Participants suggested a few solutions pertaining to the game design and marketing of educational games.

As mentioned before, enjoyment in educational games can offset their negative reputation. To design for enjoyment, participants similarly recommended certain game elements within the constructs of story, gameplay, and atmosphere. According to participants and game researchers (Brown & Cairns, 2004; Ermi & Märyä, 2005; Schell, 2014), story is the series of connected events that the player experiences (e.g., "the sequence of events that unfolds in your game" (Schell, 2014, p. 51)); gameplay is how the player's interaction with the game impacts events (e.g., "what you're doing in the game"); and atmosphere is the distinct mood of the game world, which affectively influences the player (e.g. "the calming atmosphere makes it easier to relax"). Recommended elements congruous with the gaming literature are discussed. Since thorough explanation of each element would take up so much space as to shift the aim of this paper, and is already provided by He (2017), only summaries are reported (Table 2).

**Table 2**

Game elements for enjoyment and engagement, accompanied by description and example

| Game element | | Description | Example |
| --- | --- | --- | --- |
| Story | Tone | General attitude of the text, voice, or other communication that conveys the story. Forms 3D characters; facilitates empathy and understanding. Should be familiar or relatable. | "I think because a lot of the dialogue was very conversational or more like a stream of consciousness, I could understand it. My thought process is similar and the way I think things is very similar so it was easy to follow." (Sophia, on *Grey Plague*) |
| | Detail | Should have detail or clarity about game world, especially unique points about protagonist and setting. | "I'd prefer a little more backstory.. I'm curious, I want more info about what [the protagonist's] childhood was like, and how she got her abilities." (Ava, on *Grey Plague*) |

|  | Relationships | Interpersonal relationships developed in the game's societies. Should allow development of complex relationships, moral dilemmas, and drama. | "The relationship between [the protagonist] and mom and between [the protagonist] and grandma both served as an interesting contrast.. The whole idea of [the protagonist] being an outcast because of her strange ability was interesting." (Aiden, on *Grey Plague*) |
|---|---|---|---|
|  | Fantastical reality | Common preference for reality-based stories, perhaps with fantastical twist. | "I want more realistic storylines instead of the typical post-apocalyptic future or robots taking over the world type of story; it's more relatable." (Ava, on games) |
| Game play | Clear goals and rules | Game goals and rules should be clear to players. Serves to avoid confusion or bad frustration and retain flow experience. | "Some parts of the game I couldn't really figure out what to do.. I didn't know you could interact with the people." (Evan, on *Grey Plague*) |
|  | Identity | Ability to feel present as a unique individual in-game. Facilitated by allowing player-controlled character movement and customization. | ""I like the more immersive quality – you're the person in the game; you could act out the person and character.. How about a dressable character?" (Felicia, on *Grey Plague*) |
|  | Experimentation/ exploration | Should have capability to autonomously explore game world, and experiment with game mechanics in game world. | "There were minor points in annoyance- like in the lungs, I kept on trying to go somewhere, and it was like 'no! You can't go there!'" (Felicia, on *Grey Plague*) |

|  | Agency/autonomy | Ability to make meaningful decisions and express will, uninfluenced by extrinsic rewards, in-game. | "I felt limited in my choices, like I could only hit certain options, and I could only go to certain places and talk to certain people." (Katie, on *Grey Plague*) |
|---|---|---|---|
| Atmosphere | Visuals | i.e. interface, images, animations. Fosters unique atmosphere via style and color palette. Should have clear distinctions between shapes and natural character and environment animations. | "The people created in particularly the Sims 2 and 3 games seem really stiff in gesture and don't have enough interaction, making them a lot less appealing than what's shown on the trailers." (Hannah) |
|  | Audio | i.e. music, sound effects. Should be apropos to game world and event player is experiencing, and should avoid intrusive repetition. | "The audio was good. All the sound effects were really good. It was just really repetitive; after listening to it for about twenty minutes, I started to get kind of annoyed." (Katie, on *Grey Plague*) |

Participants felt story, atmosphere, and gameplay elements were insufficient in educational games; Ava, group B and non-gamer, remarked that "educational games are boring because there's no or little storyline or characters or appealing graphics or interactivity." When playing *Grey Plague,* participants were surprised because the game was "fun," "pretty entertaining," "awesome," etc. As Hannah put it, "[*Grey Plague*] is different from other educational games because it has interaction, it has basically every aspect of a game that is necessary for deep involvement.. So yeah, I was really immersed in this game." *Grey Plague* was contrary to participants' beliefs. Participants gave effective atmosphere and story as the reason for their unexpected enjoyment. These game design suggestions support research promoting stealth learning, a concept that suffers from a lack of empirical evidence (Annetta, 2010; Paras & Bizzocchi, 2005; Shute, 2011). Stealth learning is purported to facilitate learning by placing learners in a state where their entire attention, concentration, and motivation is towards the learning technology. This state is called "flow" (Shute, 2011, p. 504), also known as immersion or intense engagement.

  Beyond changing an educational game's design, game developers can overcome barrier 1 by changing their student-targeted marketing to emphasize unique selling propositions (USPs) conducive to enjoyment, or advertise the educational game as another, more preferred game genre. Regarding USPs, story, gameplay, and genre were the most common USPs that enhanced participants' interest and perceived enjoyment in an educational game. Six participants reported story as a USP, and suggested that the best way to market story was to provide an seemingly intriguing, enjoyable blurb - "the main overview of the story, not the ending, not the conclusion, but what it's about" (Felicia, group B and gamer). Four mentioned interesting gameplay and genre as additional USPs. Genre is another solution to barrier 1 - by marketing an educational game as a more preferred, reputable game genre than educational, the educational game may get more visibility and perceived value. Participants and gaming researchers indicate that the most preferred game genres are simulation, role-playing, shooting, and adventure; however, note that genre preferences significantly vary when considering attitudinal and behavioral game-playing factors (Lee et al., 2007; Elliott, Golub, Ream, & Dunlap, 2012). In alignment with the benefits of preferred genre marketing, Samuel, group B, occasional gamer, noted that "if [*Grey Plague*] was in the gaming market like Steam, it'd probably show up in the educational category, where people don't look and aren't interested." Felicia similarly substantiated preferred genre marketing by stating that she would value *Grey Plague* more if it was "tagged as a point and click adventure rather than an educational game about learning about yourself. That doesn't sound like something I would want to spend money on, but point and click adventures – I love those." This changed perception based on genre relates to findings (Apperley, 2006; Wolf, 2001) that consumers, from sole knowledge of a game's genre, make assumptions about the game's gameplay structure and elements and perhaps, story, with some genres eliciting more positive assumptions than others. Note that changing a game's marketed genre does not signify that the game itself must become conventionalized to that genre's generic standards. Indeed, Apperley (2006) stipulates that games should preferably innovate within a genre - novelty within familiarity - to best entertain players.

*3.2.2. Barrier 2: incompatibility with older audiences*

Six participants did not feel as though current educational games were compatible with their age and maturity, and had an unfulfilled demand for educational games targeting a wider, older audience:

Researcher: What would you like to improve most about edutainment in general?

- Ava: Make them more appealing to high schoolers and mature audiences..
- Samuel: What would I improve..? Well, most educational games I see are little kid games. An improvement would be to reach an older audience.
- Evan, group A, gamer: It'd be nice to have more of them. Helpful for them to be more available to a more extensive audience. More people would be interested in it.

R: So, make it appeal to a wider audience?

E: Yep.

Participants viewed this issue as only pertinent to educational games, because when asked the same question about games ("what would you like to improve most about games in general?") participants' answers contained no mentions of audience. Participants felt that older audiences had needs that were unfulfilled since participants' past exposure to educational games consisted only of "little kid games" targeting younger demographics. This finding seems to be new to the literature. Studies on older populations (Annetta et al., 2009; Ke, 2008; Papastergiou, 2009; Watson, Mong, & Harris, 2011) do not intentionally measure their sample's demand for educational games. Schutter (2011) addresses this gap, but he discovers a demand in much older—45-85 years old—individuals. This audience demand makes sense as most educational games focus on primary to middle school students, and rarely on higher education (Giannakos, 2013; Mayer et al., 2014; Vangsnes, Økland, & Krumsvik, 2012).

*3.2.2.1. Solutions to barrier 2*

Participants suggested that older audiences desire more complexity (e.g., "I'd like to learn more complex stuff"; "the story could use more complexity"). Thus educational games for older audiences should have more complex stories, learning content, and—according to two participants—controls. This expands findings that difficulty should match the player's skill levels (Ke, 2008; Sweetser & Wyeth, 2005), seeing as optimal difficulty in an educational game must take into account learning content as well as story and gameplay.

The finding of this solution helps answer Papastergiou's (2009) inquiry to evaluate the impact of different levels of complexity within educational games. Whether or not participants' beliefs align with older audiences' actual needs is debatable, as self-reports for other populations does not always reflect reality (Kirschner, 2017).

*3.2.3. Barrier 3: incompatible difficulty*

Four participants remarked that difficulty with attending to or comprehending the content in *Grey Plague* presented barriers to engagement:

- Evan: The dialogue seemed to be too long at times. A bit more interaction.. Maybe if you can interact with something, highlight it in some way, make it different from the scenery.
- Edward: ..I found that while I was reading all that, I was starting to lose attention a couple times. So maybe, not as much talking? But still being involved?

The definition of difficulty is unpleasant exertion to accomplish something laborious, difficult to understand, or not easy to do (Nicholls & Miller, 1983). It was evident that these four did not find attending to the dialogue portions of *Grey Plague* to be easy, and had to make extra effort to do so. Since all four participants were from group A, this reasoning insinuates that game A's differentiated section, in its minimally interactive state and dry tone, did not facilitate attention and comprehension as much as game B's. Because barrier 3 mostly emerged from a question directed at *Grey Plague* specifically, this barrier is likely specific to individual educational games. This barrier is unique to educational games because the learning component adds another dimension to optimizing game difficulty.

Note that difficulty does not always pose a threat to engagement. Entertainment gaming researchers refer to the feelings induced by difficulty as 'frustration', and perceive these feelings as a mix of anger and helplessness. They consider frustration in two different dimensions: detrimental, "at-game frustration," (Gilleade & Dix, 2004, p. 230) or "pleasurable frustration"

(IJsselsteijn et al., 2007, p. 108). Pleasurable frustration enhances the player's experience and motivation, and induces catharsis and stress relief after challenges are overcome; thus, it should not be avoided (Ferguson & Olson, 2013; IJsselsteijn et al., 2007). Detrimental frustration, however, is a detriment to engagement and induced by difficulties with the controls (i.e. at-game frustration) or difficulties with comprehending the complexities of game dialogue or other game elements (i.e. in-game frustration) (Gilleade & Dix, 2004). Unlike with entertainment games, detrimental frustration with educational games seems to mostly manifest via difficulties with the learning content. Additionally, the frustration is not made of anger or helplessness; it seems to be composed more of boredom and restlessness. Seeing as the composition and causation of detrimental frustration is different for educational games, the proposed solutions to this barrier are probably distinct to educational games.

*3.2.3.1. Solutions to barrier 3*

To overcome barrier 3, participants recommended that difficulty should be personalized to the player's motor and spatial skills and intention to use (e.g., having degree of difficulty settings 'easy,' 'medium,' 'hard,' etc; having adaptive, dynamic difficulty). Additionally, due to the learning component of educational games, the learning content should suit the player's attentional control and comprehension abilities. The learning content itself need not become simplified or more complex; it was the delivery that participants mostly complained about. Seeing as this study's aim focuses specifically on educational games, only methods to optimize the difficulty of learning content are discussed, which include the following: pacing, and aids for comprehension such as voiceovers.

Pacing, also known by cognitive psychologists as segmentation (Mayer & Moreno, 2003), is the rate at which content such as learning content, story, or enemies is delivered. If too much content is delivered within a short time, the player will feel burdened and find the content difficult to comprehend, resulting in cognitive overload. Optimal pacing occurs when content is delivered in small, incremental steps, with the presented content getting progressively more complex and difficult (Khenissi et al., 2016). The benefits of tackling learning in incremental, attainable subgoals are numerous - increased self-directed learning, self-efficacy, and intrinsic interest - and have been supported through the past decades (e.g., Bandura & Schunk, 1981; Catrambone, 1998). An example of optimal pacing in-game could be having dialogue split into one or two sentence sections that must be interacted with to see the next section of the dialogue; this facilitates interaction and sequential learning. As Katie reasoned, "for this game, most of the information was condensed into one part, so if I missed that one part I wouldn't really understand it.. For information in the game.. I'd like for it to be spread out"; Hannah agreed, explaining that "that way, people could not rush through it and learn about it a little better." In line with this reasoning, Schell (2014) used "interest curves" (p. 282) to explain that interesting moments must be dispersed evenly throughout a game to retain interest, and Chang, Liang, Chou, and Lin (2017) found that excessive multimedia had negative outcomes on learning. More relevant, Mayer and Moreno (2003) found that consumption of too much content can result in cognitive overload, in other words, the cognitive processing necessary to consume the content exceeding the actual processing capacity of a learner's cognitive system. Cognitive overload causes poorer learning; namely, decreased germane attention, deep processing, and comprehension (Mayer & Moreno, 2003). Beyond easing learning, pacing also promotes reflection on what was learned. While the definition of reflection varies depending on context, it is generally known to be a mental process facilitating deep understanding of what was newly learned, in which new knowledge is evaluated for meaning and integrated with pre-existing knowledge (Ke, 2008; Moon, 2013). The importance of reflection in learning is consistently and extensively substantiated by education and educational gaming research (Moon, 2013; Habgood, 2007; Ke, 2008; Watson, Mong, & Harris, 2011).

     Aids for comprehension, as suggested by participants, include relevant interactions, visuals, and audio. The listed aids are congruous with research advocating active learning (e.g., Freeman et al., 2014) and Clark and Mayer's (2016) cognitive theory of multimedia learning. Clark and Mayer found that text and graphics pertaining to the text, rather than just solely text, are more conducive to learning. Additionally, they state that combining the two channels used to process information - auditory and visual - facilitates in-depth comprehension of the learning material. With *Grey Plague,* participants felt that the presence of interactive diagrams facilitated their learning. For instance, it was observed that a majority of participants spent about equal time mousing over a diagram and reading the accompanying text. Another observed behavior was their going back and forth between perusing diagram and text, perhaps to connect the two in forming conceptions of the content. In alignment, Ava reported that "I learned so much more because I'm a visual and kinesthetic person." Regarding audio, three participants desired voiceovers to help them comprehend. Sophia explained that "I'm an auditory learner.. I learn better if I hear it, so I might've preferred hearing someone reading the lines versus just beeping." One participant made the exception that audio appropriateness depended on the game: "you don't really need voice actors when you have a game like this with mostly dialogue." Congruous

with this consideration for context, Kirschner (2017) points out that the most optimal aid for learning comprehension actually depends more on what subject the educational game is teaching and other such contextual information rather than students' self-categorization - "pigeon-holing" (p. 167) - into well-known learning styles (e.g. visual, auditory, verbal, kinesthetic).

*3.2.4. Barrier 4: price*

Four participants mentioned that price would be a barrier to self-directed engagement, with certain contexts being the solution to this barrier:

Researcher: Would you buy the finished game [*Grey Plague*]?

- Katie: Depends on how much it is.. I don't know if I'd necessarily buy it.
- Felicia: It depends on the price. I guess I would be willing. It depends on the context.
- Samuel: Would I buy? Perhaps for a few dollars, up to five. In what context? I'd definitely if a teacher suggested it.

Barrier 4 is comparable to Shapiro et al.'s (2017) finding that the fourth most common barrier to MOOC engagement was a "lack of resources such as money" (p. 48). While it is strange that price was a similarly common barrier for the sample in Shapiro et al.'s study (aged 18 to 55 and above, mostly earning steady, taxable incomes) and this study's sample, who should have less monetary resources due to age, perhaps the reason why barrier 4 wasn't more common was because this study's participants resided in a high income area. As Levell and Oldfield (2011) from the Institute for Fiscal Studies found, high income households are more willing to spend money on leisure goods and services. An alternative or additional reason could be because younger people are more prone to compulsive buying (Dittmar, 2005).

Price was a barrier due to the insufficiency of participants' perceived enjoyment or learning usefulness with educational games; no participants attributed this barrier to lack of money. Note that enjoyment and learning have potential for overlap, e.g. for students who enjoy learning. To somewhat demarcate the two terms, for this study, perceived enjoyment signifies beliefs that a technology will cause an entertaining experience regardless if it is due to learning, while perceived learning usefulness is the beliefs that a technology will advance the user's success, or in other words, its "overall impact of system use on job performance" (Davis, 1993, p. 477). This need for learning usefulness is congruous with many previous research, including the requirements of the TAM (Davis, 1993), and more similarly, Bourgonjon et al.'s (2010) finding that learning usefulness had a "strong", positive impact on preferences for video game use in classrooms. Interestingly, unlike enjoyment, learning usefulness did not seem necessary to evoke intention to purchase an educational game. Six participants indicated they would pay for an educational game purely based on sufficient perceived enjoyment alone: at a price point of 10 to 40 USD (price point varied by participant), these six would buy *Grey Plague* because it was "fun" or "I liked it," or "I want to finish it. I really do." Likewise, Belanger and Thornton (2013) analyzed MOOC pre-course surveys and found that a majority of MOOC students signed up because they believed the taught subject was enjoyable or interesting to learn. Moreover, previous research (Ha, Yoon, & Choi, 2007; Mandryk, Atkins, & Inkpen, 2006) found that perceived enjoyment was a better predictor than perceived usefulness for acceptance of video games. But, video games do not equal educational games. In fact, despite comparisons to past research, this barrier is hard to substantiate: related, published research examines similar, but not identical contexts - the role of price in student purchases of educational games. Notwithstanding, barrier 4 and its causation corroborates this study's aforementioned findings that students demand for enjoyment in educational games. Therefore, educational games with sufficient perceived enjoyment may not face barriers 1 and 4.

*3.2.4.1. Solutions to barrier 4*

If an educational game is not perceived enjoyable enough to avoid this barrier, student purchase likelihood can be increased through teacher encouragement. Felicia gave a fitting example: "let's say I had a class, and the teacher recommended this game, or like had a good review for this game, I guess I'd give it a shot. I'd pay $5 or less." Teachers can encourage educational game purchase via three categories found by Lai (2015) when examining the impact of teacher encouragement on student self-directed use of technology: capacity support (i.e. mentioning specific technologies, guiding on their use), behavior support (i.e. using technology in-class, providing technology), and affective support (i.e. justifying and advocating for technology use). Teacher encouragement serves to enhance awareness, enhance the perceived usefulness, and improve the accessibility of educational games (Lai, 2015). Although participants only alluded to affective support as a solution to barrier 4, this is plausibly due to a lack of data rather than an insignificance of capacity and behavior support. In fact, Lai explains that the other two supports are important because affective support by itself does not equip students with the knowledge needed to use the recommended games.

This influence of teachers is supported by literature in psychology and sociology. Two theories come to mind: subjective norm and norm of obedience to authority. Although the impact of subjective norm has conflicting evidence, Bourgonjon et al. (2011) narrows it down to a more substantiated meaning: subjective norm is the tendency of a user to conform to the perceived wishes of most people important to him or her on whether to accept a new technology or not. As mentioned before, educational games seem to be new, unfamiliar technologies to students, thus subjective norm supports the importance of valued teachers' opinions on student purchase of educational games. The norm of obedience to authority, according to Smith, Mackie, and Claypool (2014), is the common view that legitimate authorities should be obeyed. Smith et al. stresses that legitimacy is a requirement for this obedience, and is established by the group consensually conferring the right to give orders onto the authority. Therefore, teachers who are valued by their students and have established legitimacy in the classroom are capable of encouraging self-engagement with educational games.

## 4. Conclusion

In conclusion, to address multiple gaps in the research on student perceptions of and engagement with educational games, this study identified four barriers and their solutions to self-directed engagement with educational games. In the process, a framework of educational game elements for enjoyment and engagement emerged (Table 2), and the relationship between engagement and performance was found insignificant. Insignificance may be due to multiple reasons, some being certain limitations of this study, such as insufficient differences between games A and B.

Barriers, from most common to least common, were as follows: (1) moderately negative reputation of educational games, (2) incompatibility with older audiences, (3) incompatibility between difficulty and skills, (4) price. Each barrier is unique to educational games. Barrier 1 and 2 were described by participants only when referring to educational games as a whole; barrier 3 addresses difficulty in playing certain educational games, complicated by the unique learning aspect of educational games; and barrier 4 was affected by perceived enjoyment and learning usefulness, which could be increased per teacher encouragement.

Several implications of this study's insights were found. Firstly, given the results, this study's findings are in alignment with researchers' (Bourgonjon et al., 2010; Muspratt & Freebody, 2007) claims that students' engagement with instructional technology is incredibly complex. A combination of efforts from teachers and developers can encourage engagement, and further research can help illuminate effective directions. Despite assurance that educational games will be straightforwardly accepted by students (Luckin et al. 2009; Palfrey & Gasser, 2013; Tapscott, 2009), this does not seem to be the case. The engagement process between students and educational games warrants more investigation.

Secondly, despite barriers 1 and 2 being the most common barriers, they were novel findings to the educational gaming literature. If findings are not false, the negative reputation of educational games should be overcome to maximize student engagement, self-directed or not. Additionally, older audiences have unfulfilled, untapped needs. To capitalize upon these needs and facilitate older students' learning, researchers and developers should listen to older audiences and adjust the marketing and design of educational games accordingly.

Thirdly, although some suggestions for overcoming barriers are similar to those found in literature describing effective game elements for learning (Annetta, 2010; Guillén-Nieto & Aleson-Carbonell, 2012), more suggestions are in alignment with effective game elements for enjoyment (Pinelle, Wong, & Stach, 2008; Schell, 2014). Additionally, a couple participants noted their engagement was more concentrated on the entertaining rather than learning content. Moreover, all participants desired for "entertainment value" in their entertainment media, including games. These findings implicate that students, controlled for confounding attitudinal factors, desire being entertained over learning in an educational game. Thus, self-directed engagement in an educational game is further encouraged by increased entertainment and does not necessarily ensure self-directed learning.

This study provides valuable insights and opens new avenues for future work. Similar research that includes larger samples, greater differences between intervention games, more thorough testing, and measurement of perceptions before intervention will better ascertain this study's findings. For future research topics, it would be desirable to investigate the reputation of educational games and older audience needs for educational games. Both these topics are insufficiently addressed by educational game developers and researchers alike and, if addressed effectively, may increase the impact of educational games.

*4.1. Limitations*

This multiple case study examines a small sample of participants who, despite proportional diversity in gender, race, academic achievement, and gaming habits, were from the same school in a high-income county. Regarding the interventions, both were short-term, and the differences between them constituted at most a quarter of total playtime and did not seem statistically significant, thus might've impacted quantitative analyses. Regarding data collection, demographic, engagement, and interview data was self-reported, which can be unreliable due to probability that participants are unable or unwilling to report actuality. Additionally, in hindsight, the performance assessment questionnaire was not optimal due to its limited number of questions and multiple choice nature, which allowed falsely correct answers via chance.

Caution should be exercised when generalizing findings to other instructional technologies and applying findings to populations with different demographics.

Appendix A

After air leaves the trachea, it goes through the:

Bronchioles

Bronchi

Alveoli

Capillaries

Gaseous exchange takes place in the lungs in the:

Bronchioles

Bronchi

Alveoli

Cardiac muscle is:

found everywhere

located in the abdomen

unique to the heart

(Select all that apply.) Select all of the following that are symptom(s) of tuberculosis.

Cavitation

Granulomas

Nodosomes

Clear sputum

Tuberculosis is transmitted through:

Infected water

Infected hands

Infected blood

Infected air

Which is most commonly collected to diagnose respiratory infections?

Saliva

Breath

Sputum

Any of the above

None of the above

What do Mycobacterium tuberculosis have that prevent their destruction by the immune system?

Cord factor

Exotoxins

Special protein

Capsule

Endotoxin

Which of the following is NOT one of the lobes of the cerebral hemisphere?

Ethmoid

Frontal

Occipital

Temporal

The central nervous system consists only of the brain.

True

False

Which of the following statements best describes homeostasis?

Keeping the body in a fixed and unaltered state

Maintaining a balanced internal environment

Altering the external environment to accommodate the body's needs

Appendix B

1. How have your opinions of Anatomy and Physiology changed after playing the game? For example, a changed interest in the topic? Any changed beliefs?
2. After playing the game, would you like to get involved more with Anatomy & Physiology, for example learn more about anatomy or medical sciences in your free time, and why or why not?
3. How much did you trust the information about anatomy given in the game?
4. Did you think Grey Plague was a real disease?
5. How do you feel about learning about anatomy in a game?
6. What do you think you learned from the game?
7. What problems with the game did you run into?
8. Did you get confused/annoyed by the controls at any point in time?
9. Would you buy the finished game at your chosen price? **If no,** why not? **If yes,** why, and how much would you pay for it?
10. During the first few minutes of playing, what was your initial opinion of this game?
11. After playing the game, did your opinion change, and if so, how?
12. How do you learn about games? E.g. through social media, news, word of mouth, etc?
13. When sharing a game to others, what do you use to share? E.g. email, facebook, twitter, etc?
14. When learning about a game, what information do you like to see provided? Information about the story, or about the unique features, etc?
15. How entertaining was the game?
16. How were the visuals? How could it be improved?

17. The story? How could it be improved?

18. The gameplay? How could it be improved?

19. The audio??

20. What aspect of games are most important to you—visuals, story, gameplay, etc?

21. What are the things that you would like to improve in this game?

22. What do you like most about this game?

23. What would you most like to improve/change about edutainment in general?

24. What would you most like to improve/change about games in general?